\documentclass{webofc}
\usepackage[varg]{txfonts}   
\begin{document}
\title{Simulation of neutron production in hadron-nucleus and nucleus-nucleus
 interactions in Geant4}
%
%

\author{\firstname{Aida} \lastname{Galoyan}\inst{1}\fnsep\thanks{\email{galoyan@lxpub01.jinr.ru}} \and
        \firstname{Alberto} \lastname{Ribon}\inst{2} \and
        \firstname{Vladimir} \lastname{Uzhinsky}\inst{1,2}
        }

\institute{JINR, Dubna, Russia 
\and
           CERN, Geneva, Switzerland
}

\abstract{%
  Studying  experimental data obtained at ITEP \cite{ITEP}
on neutron production in interactions of protons with
various nuclei in the energy range from 747 MeV up to 8.1 GeV, we have found  that slow neutron
spectra have scaling and asymptotic properties \cite{jetp}. 
 The spectra  weakly depend on the collision energy at momenta of projectile 
 protons larger than 5 -- 6 GeV/c. 
These properties are taken into account in the Geant4  Fritiof (FTF) model.
The improved FTF model describes as well as the Geant4 Bertini model 
the experimental data on neutron production by 1.2 GeV and 1.6 GeV
protons on  targets (Fe, Pb) \cite{Leray} and  by 3.0 GeV 
protons on various targets (Al, Fe, Pb) \cite{ISHIBA}.
For neutron production in antiproton-nucleus interactions, it was demonstated that
the FTF results are in a satisfactory agreement with experimental data of the LEAR 
collaboration \cite{LEAR}.
The FTF model  gives  promising results for neutron 
production in nucleus - nucleus interactions at projectile energy 1 -- 2 GeV per 
nucleon \cite{Yurevich}.
The observed properties allow one to predict neutron yields in the nucleus-nucleus interactions
at high and very high energies. Predictions for the NICA/MPD experiment at JINR are presented.
}
\maketitle
\section{Introduction}
\label{intro}
For Beam-Energy Scan program running now at RHIC (BES), and new experiments at 
future accelerator complexes -- such as FAIR at GSI and NICA at JINR -- it is very 
important to know the yield of neutrons for detector design, estimations 
of radiation doses, creation of Zero Degree  and hadron
calorimeters. For these reasons, it is needed to verify and develop Monte Carlo 
generators for neutron production in hadron-nucleus and nucleus-nucleus 
interactions. There are some generators -- UrQMD \cite{URQMD1,URQMD2},  
LAQGSM \cite{LAQGSM1}, Fluka\cite{Fluka1}  
and Geant4 \cite{Geant4} models, which are able to simulate nucleus-nucleus interactions. 

The standard UrQMD model does not treat the nuclear remnant.
 Because the source codes of LAQGSM and Fluka are not publicly available,
 we use and develop Geant4 FTF generator, whose source code is open. 
We also apply Geant4 Bertini and  Binary Cascade models for calculations 
and comparisons with the FTF results and experimental data.  
The Geant4 FTF model is based on the Fritiof model \cite{Fritiof1,Fritiof2} of
 the  LUND university. Description of the Geant4 FTF model is given in the paper \cite{Geant4}.

We have performed an analysis of a large set of ITEP experimental data \cite{ITEP} 
on neutron production  with energies from 7.5 up to 190 MeV in proton-nucleus 
interactions. 
Asymptotical and scaling properties of produced neutrons have been found.  
We used these properties of neutron spectra to develop the Geant4 FTF model. 
Then, the experimental data on neutron production  
for nuclear reactions induced by protons at 1.2, 1.6 GeV \cite{Leray}, 
and 3.0 GeV \cite{ISHIBA} for various targets have been considered.  
We have tried to describe these experimental data using the 
improved FTF, Bertini  and Binary cascade models.
Experimental data on neutron production in anti-proton -- nucleus interactions of the LEAR 
collaboration \cite{LEAR} have been also studied. 
We applied  calculations with FTF+Preco (FTFP) and FTF+Binary cascade (FTFB) models
for comparison with the LEAR experimental data.
For the FTF model verification we used also
experimental data on neutron production in collisions of light nuclei with nuclei 
at 1 and 2 GeV per nucleon.  
The obtained results allow us to give FTF model  
predictions for the future experiment MPD at the new accelerator NICA at JINR. 

\section{Main assumptions and development of the FTF model}
The main assumptions of the Fritiof model are quite simple.  
It is assumed that  hadrons turn into excited states in hadron-hadron collisions.
If only one hadron is excited, the process is called diffraction dissociation. 
Excited states of hadrons are characterized by their respective masses. Minimal masses of excited 
states are different in various models. 
In the original Fritiof, the minimal mass is 1.2 GeV, in UrQMD is 1.46 GeV, 
and in Hijing is 2 GeV. 
 In the Geant4 FTF model, we decreased the minimal mass of excited states to 1.16 GeV.
 This reflects on  simulation of events
with small multiplicity, in particular, in description of the diffraction dissociation.
The excited states of hadrons are considered as quark-gluon strings.
The Lund fragmentation model is used for quark-gluon string fragmentations.
Neutrons are produced at the fragmentation of the quark-gluon strings,
 in the diffraction dissociation and  in proton charge-exchange processes.

For simulation of hadron-nucleus and nucleus-nucleus interactions,
the  Glauber approach is applied in the FTF model to determine the 
number of nucleons participating in inelastic interactions.  After that, string 
creation and fragmentation are simulated. 
The Glauber approach implemented in FTF is not sufficient for description of the observed destruction 
of nuclei. 
Thus, a reggeon cascading model \cite{reggeon} of nuclear destruction is applied in the FTF model,
which plays a very important role in the production of nucleons. 
 The model, inspired by Reggeon theory, assumes that  spectator nucleons can be involved by the
 participating nucleons at the first stage of the interaction. 
A probability to involve a spectator nucleon having coordinates in the impact 
parameter plane $\vec{b_{i}}$ by a participating nucleon with coordinates 
$\vec{b_{j}}$ is written as:

\begin{equation}
P_{ij}=C_{nd}e^{-(\vec{b_{i}}-\vec{b_{j}})^{2}/R^{2}},\ R \simeq 1.2 \ (fm) 
\label{myeq1}
\end{equation}
 $C_{nd}$ is the parameter of nuclear destruction, given by the formula: 
\begin{equation}
C_{nd}=0.0048 \cdot A \cdot e^{4(y_{p}-2.1)}/(1+e^{4(y_{p}-2.1)}),
\label{myeq2}
\end{equation}
where A is the mass number of the target nucleus and $y_{p}$ is the rapidity 
of the projectile in the target nucleus rest frame.

In the FTF model, momenta of participating and involved nucleons are sampled 
according to the expression: 
\begin{equation}
P(p_{z},\vec{p_{T}})\propto e^{-|\vec{p_{T}}|^{2}/\left<|\vec{p_{T}}^{2}\right>} 
\cdot e^{-(x^{-}-1/N)^{2}/(0.3/N)^{2}},
\label{myeq3}
\end{equation}
$$x^{-}=(E-p_{z})/(E_{N}-P_{N}),$$
\begin{equation}
\left<|\vec{p_{T}}^{2}\right> = 0.035+0.04\cdot e^{4(y{_p}-2.5)}/(1+e^{4(y{_p}-2.5)}),
\label{myeq4}
\end{equation}
where  $N$ is  multiplicity of the involved  and participating nucleons, 
$x^{-}$ is the light-cone momentum fraction, $E_{N}$ and $P_{N}$ are the total 
energy and the momentum of all interacting nucleons of the target nucleus. 
Numbers in Eqs. (1)-(4) were determined \cite{jetp} in order to reproduce the experimental data \cite{ITEP}.  

\section{Analysis of experimental data on neutron production}

Studying the experimental data on neutron production in the interactions of protons at 
 energies  from 747 MeV up to 8.1 GeV with
various targets  \cite{ITEP}, we noted  \cite{jetp} 
that  neutron spectra have interesting regularities.  
We found that slow and fast neutron production cross section
  are weakly energy-dependent at $P_{lab} > $5 -- 6  GeV/c at neutron emission angle 119$^0$.
In other words, neutron spectra approach the asymptotic regime at
$P_{lab} > 5$ GeV/c.
\begin{figure}[h]
\centering
\includegraphics[width=7cm,clip]{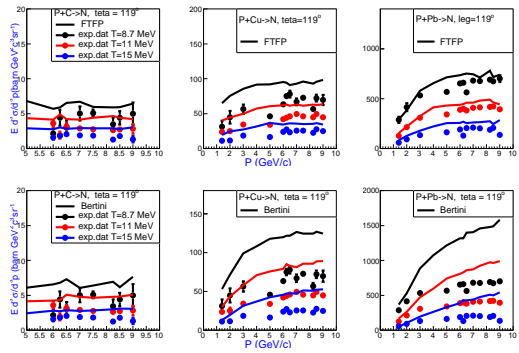}
\caption{Inclusive cross sections of neutrons produced in pC, pCu, pPb interactions at angle 119 degrees
as functions of projectile momenta.
Points are experimental data \cite{ITEP}. Lines are calculations by the FTFP model in upper figures, and  
the Bertini model in bottom figures. Black points and lines correspond to neutrons with energy 8.7 MeV, 
red points and lines correspond to neutrons with energy 11 MeV, blue points and lines correspond to neutrons 
with energy 15 MeV.}
\label{fig-1}       
\end{figure}
\vspace{-5mm}

Let us consider  the energy and target 
mass dependencies of the slow neutron  cross sections at neutron kinetic energies T = 8.5, 11, 
and 15 MeV. We believe that at these energies the data represent the main properties of 
the slow neutron spectra. The corresponding experimental data are presented in Fig.1
for proton interactions with C, Cu, and Pb targets. 
Calculations by the FTF and Bertini models of the Geant4 are also shown in Fig. 1.
 As seen, for the lead and copper nuclei, the asymptotic 
behaviour takes place at $P_{lab}\sim $ 5 GeV/c. 
Because the experimental data for the carbon nuclei start at projectile  momenta 6 GeV/c,
only asymphtotic regime for the carbon nuclei is seen at $P_{lab} \geq $ 6 GeV/c.
The FTF results for p+C and p+Cu interactions are closer to the experimental data 
than the Bertini results. 
The Bertini model produces too many slow neutrons in the backward hemisphere at 119 degrees.     
For the lead target, the FTF model describes the experimental data better than the Bertini model too.
According to the nuclear scaling hypothesis, the inclusive cross sections of nucleons scale 
as $A^{n}$, where $A$ is the mass number of target. The analysis of the experimental data \cite{ITEP} 
demonstrates  \cite{jetp}  that the scaling coefficient of the evaporated neutron inclusive 
cross sections depends on the kinetic energy and can be presented as $A^{n} f (E)$.

Let us consider the experimental data on neutron production 
in proton interactions with Al, Fe, Pb at energy of projectile protons of 3 GeV, presented in
 paper \cite{ISHIBA}. In Fig.2, inclusive cross sections of neutrons produced in p+Fe interactions 
  are shown at various angles of emitted neutrons: 15, 30, 60, 90, 120 and 150 degrees. 
\newpage
\begin{figure}[h]
\centering
\includegraphics[width=7cm,clip]{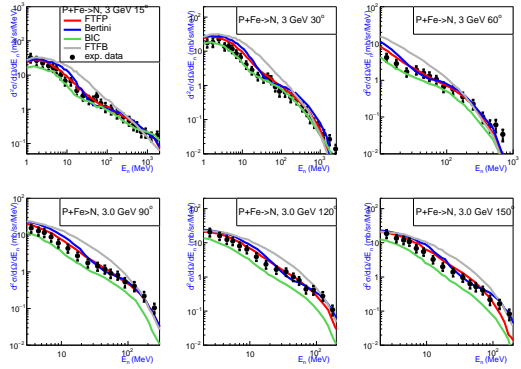}
\caption{Inclusive cross sections of neutrons produced in p+Fe interactions at projectile momentum 3 GeV/c 
at various emission angles of neutrons.
Points are experimental data \cite{ISHIBA}. Lines are calculations by the FTFP, FTFB, Bertini and
 Binary cascade models.}
\label{fig-2}       
\end{figure}
\vspace{-5mm}

The FTF+Preco (FTFP), Bertini, Binary cascade model (BIC) and FTF+Binary cascade model (FTFB) 
calculations are also given. 
Preco and Binary Cascade models were added to the FTF model for the simulation of secondary slow particles
propagation inside the target
nucleus. The Preco model considers  absorptions of the secondary slow particles. 
The Binary Cascade model takes into account all cascading processes and absorbtions of 
the secondary slow particles. Due to this, FTFP is faster than FTFB.
 As seen in Fig.2, the FTFB calculations (gray lines) are above the experimental data.  
The Binary Cascade model calculations (green lines) underestimate neutron production at large angles. 
The FTFP (red lines), Bertini (blue lines) model calculations are close to each other and 
describe the experimental data quite well. 
The same good agreement between the experimental data and  FTFP as well as Bertini model calculations are 
obtained for p+Al interactions. 
When we consider neutron spectra in the interactions of proton with lead nuclei at energy 3 GeV,
 all four models -- FTFP, FTFB, Bertini and Binary cascade -- show quite good agreement with the experimetal data.
We can conclude that the FTFP model works well for proton interactions with light and heavy nuclei at energy 3 GeV
and higher (see above). 

We also tested the FTFP model at lower energies. 
For this, the experimental data on neutron production by 1.2 and 1.6 GeV protons on Fe and Pb nuclei
presented in the paper \cite{Leray} were considered.  
In Fig.3, we present the double differential cross sections of neutron production in 
p+Pb interactions as  functions of neutron kinetic energy at various angles of emitted neutrons,
 from 0 to 160 degrees.   
 We compare the experimental data with results of the FTFP, FTFB, Bertini and Binary cascade models.
At larger angles, FTFP overestimates the neutron spectra at energy larger than 40 MeV. 
The Binary cascade model underestimates production of neutrons in backward hemisphere and overestimates 
experimetal data at small angles 0, 10 degrees. 
The Bertini model calculations are quite close to the experimental data.
 At small and intermediate angles, the FTFP describes the most part of the experimental data. 
 The FTFB results are worse than the FTFP results. FTFB underestimates neutron yield at small angles less 
than 30 degrees and overestimates the production of energetic neutrons for angles more than 90 degrees.
We obtained  similar results for neutron spectra for lead target at the energy 1.2 GeV
and for iron target at the energies 1.2 and 1.6 GeV.  
On the whole, we can conclude that the FTFP model describes satisfactorily neutron spectra in 
proton-nucleus interactions at initial energy more than 1 GeV.
\newpage
\begin{figure}[h]
\centering
\includegraphics[width=7cm,clip]{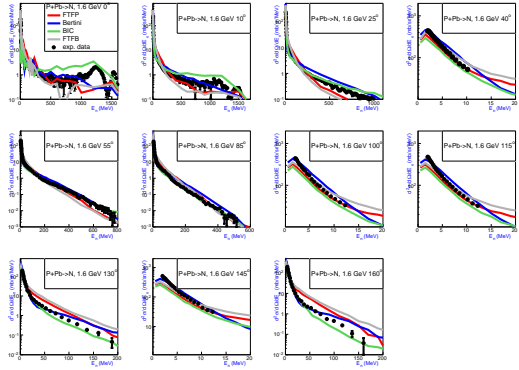}
\caption{Inclusive cross sections of neutrons produced in p+Pb interactions at projectile energy 1.6 GeV 
at various emission angles of neutrons.
Points are the experimental data \cite{ISHIBA}. Lines are calculations by the FTFP, FTFB, Bertini,
 Binary cascade models.}
\label{fig-3}       
\end{figure}
\vspace{-5mm}
Let us turn to neutron production in antiproton-nucleus interactions.
For the simulation of antiproton-proton interactions, the FTF
model uses the main assumptions of the Quark-Gluon-String Model \cite{DPM}.
The FTF model assumes production and fragmentation of quark-anti-quark and
diquark-anti-diquark strings in the mentioned interactions. The main ingredients of
the FTF model are cross sections of string creation processes
\cite{Baldin} and application of the LUND string fragmentation algorithm.
In antiproton-nucleus interactions, neutrons can be produced at 
diquark-anti-diquark string fragmentation,
and in the  processes which we mentioned for proton-nucleus interactions.
In  Fig.4, we show kinetic energy distributions of neutrons produced in 
anti-proton interactions with heavy nuclei -- Ta and U, at projectile momentum 1.22 GeV/c.
We compared the experimental data \cite{LEAR} with calculations by the FTFP and FTFB models.  
\begin{figure}[h]
\centering
\includegraphics[width=6.5cm,clip]{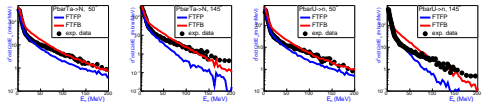}
\caption{Inclusive cross sections of neutrons produced in antiproton interactions 
with Ta and U at projectile momentum 1.22 GeV/c.
Points are experimental data \cite{LEAR}. Blue and red lines are calculations by the FTFP and FTFB models, 
correspondingly.}
\label{fig-4}       
\end{figure}
\vspace{-5mm}

The experimental data are given for forward (50 degrees), and backward (145 degrees) angles. 
The FTFP describes slow neutron production at energy lower than 40 MeV and 
underestimates fast neutron production.
The FTFB results are in reasonable agreement with the experimental data at forward
angles. The FTFB model slightly underestimates the production of fast neutrons at backward angles.
On the whole, the FTFB model better describes  neutron spectra in antiproton-nucleus interactions
than FTFP.
A correct description of neutron experimental data in the FTF model is 
very important for the estimation of radiation doses for the future PANDA experiment.

Another set of data, which we used for the FTF model validation, 
are on neutron production in nucleus-nucleus interaction at 
energies  1--2 GeV/c/nucleon.  The experimental  data were obtained at 
JINR (Dubna) \cite{Yurevich}.
We compared the FTF model calculations
with the experimental data on kinetic energy distributions of neutrons produced in
p, d, He-4, C + Pb interactions at different angles of emitted neutrons.
We obtained  that the FTF calculations are in a good agreement with the experimental data for 
p+Pb and d+Pb  interactions.
For C+Pb interactions, the FTF model gives larger multiplicity of fast neutrons 
than the experimental data.
It is also demonstrated in \cite{jetp} that FTFP reproduces spectra of neutrons 
for carbon interactions with light nuclei (C, Al).
 We consider the obtained results  as promising ones. 
  
\begin{figure}[h]
\centering
\includegraphics[width=6cm,clip]{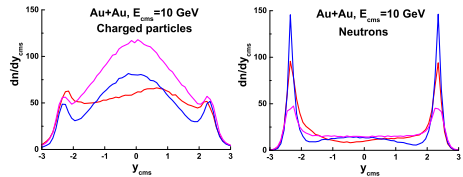}
\caption{Rapidity distribution of  charged particles (left)  and neutrons (right)
produced in Au+Au interactions at NN center of mass energy  10 GeV. Blue, red, violet lines are calculations by the 
FTFP, QGSM (Shield version), UrQMD 1.3, corrrespondingly.}  
\label{fig-5}       
\end{figure}
\vspace{-5mm}

The FTFP model predictions for the NICA/MPD experiment at JINR are presented in Fig. 5.
We simulated with the FTFP, UrQMD 1.3,  QGSM (SHIELD version)  models  Au+Au interactions 
at center of mass energy  of NN collisions  10 GeV.
Charged particle and neutron rapidity distributions are plotted in Fig. 5.
Large differences between the model calculations are observed for neutron spectra.
FTFP predicts multiplicity of neutrons produced in the fragmentation regions higher
than UrQMD and QGSM.
This difference complicates the design of a forward hadron calorimeter which can be 
used for centrality determination. The FTF estimation is more realistic than UrQMD and QGSM ones, 
because we checked the FTF results
for neutron spectra in the hadron-nucleus and nucleus-nucleus interactions. 

\section*{Acknowledgements}
A. Galoyan is thankful to heterogeneous computing team of LIT
JINR (HybriLIT) for support of calculations  and Professor Cheuk-Yin Wong for
helpful discussions.
 
%
%
%

\end{document}